\documentstyle[12pt]{article} \begin{document}
\def\<{\langle}
\def\>{\rangle}
\newcommand{\EQ}{\begin{equation}}
\newcommand{\EN}{\end{equation}}
\newcommand{\EQA}{\begin{eqnarray}}
\newcommand{\EQN}{\end{eqnarray}}
\newcommand{\EQAN}{\begin{eqnarray*}}
\newcommand{\EQNN}{\end{eqnarray*}}
\newcommand{\e}{{\rm e}}
\newcommand{\Sp}{{\rm Sp}}
\newcommand{\Tr}{{\rm Tr}}
\newcommand{\p}{\partial}
\newcommand{\h}{{1\over 2}}
\newcommand{\tright}{$\triangleright \quad$}
\newcommand{\tleft}{$\triangleleft \quad$}
\newcommand{\IR}{\relax{\rm I\kern-.18em R}}
\renewcommand{\theequation}{\arabic{equation}}

\begin{titlepage}
\begin{flushright}
  hep-th/9909022\\
\end{flushright}
\vspace{1cm}
\begin{center}
{\bf M-THEORY AND DEFORMATION QUANTIZATION} \\
\vspace{1cm}
{\bf D. Minic} \footnote{e-mail: minic@physics.usc.edu} \\
\vspace{.5cm}
Department of
Physics and Astronomy, University of Southern California, Los Angeles, CA 90089-0484 \\
\vspace{.5cm}
{\bf Abstract}
\end{center}

We discuss deformation quantization of the covariant, 
light-cone and conformal gauge-fixed p-brane actions ($p >1$) 
which are closely related to the structure of the 
classical and quantum Nambu brackets. 
It is known that deformation quantization of the Nambu bracket 
is not of the usual Moyal type. Yet the Nambu bracket 
can be quantized using the Zariski deformation 
quantization (discovered by Dito, Flato, Sternheimer and 
Takhtajan) which is based on factorization of polynomials in several real 
variables.
We discuss a particular application of the Zariski deformed 
quantization in M-theory by considering the problem of 
a covariant formulation of Matrix theory. We propose  
that the problem of a covariant formulation of Matrix Theory can be solved using
the formalism of Zariski deformation quantization of the triple Nambu
bracket.

\end{titlepage}

\section{Introduction}

Even though a fully background independent non-perturbative 
formulation of M/string theory is still
largely mysterious, some notable progress 
toward a background dependent 
formulation of M/string theory has been 
recently made. At the moment, perhaps the most promising background
dependent formulations of M/string theory are Matrix theory \cite{matrix}
and AdS/CFT duality \cite{ads}. The two approaches appear to be intimately 
related, as pointed out in \cite{pol}.

Both formulations suggest a duality between maximally supersymmetric
Yang-Mills theory (or appropriate conformally invariant field theory) 
and supergravity in a particular background. 
However, the underlying physical reason for such a duality is
not fully understood. The flat space-time limit of this duality appears 
particularly puzzling at present.

It has also been recently uncovered \cite{nonc} that string 
and light-cone M-theory in the background of $B$ fields, taken 
in appropriate limits,  are described by supersymmetric 
Yang-Mills theory on a non-commutative space. 
This non-commutative Yang-Mills theory is formulated using the 
Moyal bracket, a structure essential in the 
deformation quantization of the Poisson brackets. 

In this article we suggest that a different kind 
of deformation quantization - called Zariski deformation quantization
\cite{flato} - is the relevant mathematical structure for the formulation of
a covariant Matrix theory.

We motivate our presentation by a discussion of deformation quantization 
of the covariant, light-cone and conformal gauge-fixed p-brane 
actions \cite{volume}, \cite{membrane}, \cite{area}, 
\cite{bars} which are closely related to the structure of the 
classical and quantum Nambu brackets \cite{nambu}, \cite{flato1}, \cite{fi}. 
It turns out that deformation quantization of Nambu brackets is {\it not}
of the usual Moyal
type. As shown by Dito, Flato, Sternheimer and Takhtajan \cite{flato} 
the Nambu bracket can be quantized using the 
Zariski deformation quantization which is based on 
factorization of polynomials in several real variables. 

In view of this fact we discuss the application of the Zariski deformed
quantization 
to the problem of a covariant formulation of Matrix theory.

The paper is organized as follows:
first, in section 2. we briefly review the connection between Nambu brackets and 
the covariant, light-cone and conformal gauge-fixed p-brane actions. Then in
section 3. 
we discuss the fundamental properties of the classical Nambu bracket and 
the Zariski deformation quantization of the same.  Then in section 4. we apply 
the results of section 3. to the problem of covariantization of Matrix theory 
in terms of the deformed Zariski product. In this section we state a precise 
proposal toward
a covariant formulation of Matrix theory.
In section 5. we compare the construction of quantum Nambu brackets in terms 
of square and cubic matrices \cite{almy} to the Zariski deformation quantization 
scheme and discuss a natural mathematical formulation of the space-time uncertainty 
principle of M-theory \cite{stu} in view of our proposal.

\section{$p$-Branes and
the Nambu Bracket}

In this section we review the relation between the various forms of the bosonic
p-brane actions
($p>1$) 
and the classical
Nambu bracket. 
Let us start from the familiar covariant bosonic p-brane action 
\cite{volume}
\EQ
S = - \int d^{p+1} \xi \sqrt{-g},
\EN
where $g = det g_{ij}$ and the induced world-volume metric $g_{ij}$ is 
\EQ
g_{ij} = \partial_{i} x^{\mu} \partial_{j} x^{\nu} \eta_{\mu \nu}.
\EN
Here $x^{\mu} (\mu= 0,1,...,d) $ 
denote the target space coordinates of a $d+1$ dimensional
bosonic p-brane. $x^{\mu}$'s are functions of
$p+1$
world-volume coordinates $\xi^{i}, i=0,1,..,p$.
The equation of motion for $x^{\mu}$
follows from (1):
\EQ
\partial_{i}(\sqrt{-g} g^{ij} \partial_{j} x^{\mu}) =0.
\EN 
The covariant p-brane action is reparametrization invariant 
\EQ
\delta x^{\mu} = \epsilon^{i} \partial_{i} x^{\mu}.
\EN
The world-volume reparametrization invariance 
allows us to choose different gauges: we discuss the light-cone and conformal 
gauges in what follows, because of their relation to the structure of the 
classical Nambu bracket.

\subsection{Volume Preserving Diffeormorphisms and
the Classical Nambu Bracket}

The light-cone gauge variables are denoted by
$x^a$ ($a=1,...,d-1$) . The light-cone
coordinates are defined as 
\EQ
x^{\pm} = {1 \over \sqrt{2}} ( x^d \pm x^0),
\EN
where in the light-cone gauge     
\EQ
\partial_{i} x^{+} \equiv p^{+} \delta_{i0}. 
\EN   
The world volume coordinates $\xi^{i}$ split as
\EQ
(\xi^{0}, \xi^{s}) \rightarrow (t, \xi^{s}).  
\EN
The light-cone bosonic p-brane action is \cite{volume}     
\EQ
S_{lc}= 1/2 \int d^{p+1} \xi ( (D_0 x^{a})^{2} - h ) , 
\EN
where $h \equiv \det g_{rs}$ ($r,s=1,2,..,p$) is the 
determinant of the induced  p-dimensional
metric and the covariant derivative
\EQ
D_0 x^{a} = (\partial_0 +u^{s} \partial_s)x^{a}.
\EN
Note that the equations of motion imply 
\EQ
\partial_s u^{s} =0.
\EN
Also, $h \equiv \det g_{rs}$ ($r,s=1,2,..,p$) 
is  determined by the following expression    
\EQ
h = (1/{p!}) \{x^{a_1}, x^{a_2},...,x^{a_p}\}^{2} .
\EN
The symbol $\{f,g,...w\}$ denotes
the classical Nambu bracket \cite{nambu}  with respect to
$\xi_{s}$   
\EQ
\{ f,g,...,w \} \equiv \epsilon^{ij..k}\partial_{\xi^{i}} 
f \partial_{\xi^{j}}g...\partial_{\xi^{k}}w ,
\EN 
where $f,g,...w$ denote functions of  $\xi_{s}$. 
Thus the light-cone bosonic p-brane action can be rewritten as follows     
\EQ
S_{lc}= 1/2 \int d^{p+1} \xi ( (D_0 x^{a})^{2} - 
(1/{p!}) \{x^{a_1}, x^{a_2},...,x^{a_p}\}^{2} ) . 
\EN
The 
original world-volume diffeomorphisms $\delta x^{\mu} = \epsilon^{i}
\partial_{i} x^{\mu}$ reduce in the light-cone gauge to the p-dimensional volume
preserving
diffeomorphisms described by the action of the
classical Nambu bracket (see section 3. for details)
\EQ
\delta x^{a} = \{ u,v,... , x^{a} \}.
\EN 
The longitudinal
coordinate $x^{-}$ is determined 
from the primary constraint $p^{a} \partial_{r} x^{a} + p^{+}
\partial_{r} x^{-} \approx 0$ and the requirement, 
implied by the equation of motion, that the
longitudinal momentum is time-independent  
$\partial_{0} p_{+} =0$.

\subsection{Conformal Diffeomorphisms}

The conformal gauge is defined as follows \cite{bars}
\EQ
g_{00} =-h, \quad g_{0a} =0, \quad g_{ab}= h_{ab}.
\EN
If $a=1$ we get the usual conformal gauge of perturbative string theory.

The parameters $\epsilon^0$ and $\epsilon^a$ from the equation
$\delta x^{\mu} = \epsilon^{i} \partial_{i} x^{\mu}$
satisfy the following relations
\EQ
\partial_0 \epsilon^0 = \partial_a \epsilon^a, \quad \partial_0 \epsilon^a =
h h^{ab} \partial_b \epsilon^0.
\EN
The conformal gauge action is \cite{bars}
\EQ
S_c= 1/2 \int d^{p+1} \xi ( (\partial_0 x^{\mu})^{2} - 
(1/{p!}) \{ x^{\mu_{1}},..., x^{\mu_{p}} \}^{2}).
\EN
Note the similarity between the structures of $S_{lc}$ and $S_{c}$. 
The obvious difference is that the form of $S_c$ is covariant with respect 
to target space indices, unlike the form of 
$S_{lc}$.

This action is invariant under
$\delta x^{\mu} = \epsilon^{i} \partial_{i} x^{\mu}$
with $\epsilon^0$ and $\epsilon^a$
given above.
One can also check that the constraints
\EQA
T_{00} = 1/2(\partial_0 x^{\mu})^{2} + 
1/(2p!) \{ x^{\mu_{1}},..., x^{\mu_{p}} \}^{2} \approx 0 \cr
T_{0a} = \partial_0 x^{\mu} \partial_a x^{\mu} \approx 0
\EQN
transform into each other under this residual symmetry \cite{bars}.

\subsection{Discretization of p-Branes and the Quantum Nambu Bracket}

Given the expressions for  $S$, $S_{lc}$ and $S_c$ there obviously exists a
natural 
relation between the structure of the classical Nambu bracket 
and the covariant, light-cone and conformal gauge-fixed p-brane actions. Thus, by 
quantizing the p-dimensional Nambu bracket we can naturally discretize either 
covariant or light-cone or conformal
p-brane action. 

What do we mean by "discretize"?
Recall that the light-cone membrane action (p=2) can be regularized 
(or discretized) by applying
the Goldstone-Hoppe map  between representation theories of the
algebra of the area preserving diffeomorphisms (APD) and 
the $N = \infty$ limit
of Lie algebras \cite{membrane}, \cite{area}.
In particular, the Goldstone-Hoppe prescription instructs us to perform
the following translation of the transverse spatial
coordinates   
\EQ
\{x^{a}, x^{b}\}  \rightarrow [X^{a}, X^{b}], 
\EN
where now $X^{a}$ denote large $N \times N$ hermitian
matrices.
Moreover, integration is translated into tracing
\EQ
\int d \xi^{1} d \xi^{2}...  \rightarrow Tr .
\EN
This dictionary translates the light-cone membrane action \cite{membrane}, \cite{area}
into the action of Matrix theory \cite{matrix}.

The obvious generalization of this prescription for the case of p-branes is
\cite{almy},
\cite{hoppe}
\EQ
\{x^{b_1}, x^{b_2},...,x^{b_{l}}\}  \rightarrow [X^{b_1},
X^{b_2},...,X^{b_{l}}], 
\EN
where for the moment the nature of $X^{b_k}$ is not precisely defined (see
sections 3. and 4.). Note that in this prescription index $b$ is either the
transverse space index $a$ in the light-cone gauge, or the
covariant
index $\mu$ in the conformal gauge; in both cases $l=p$. The object on the
right-hand side is what we call 
the quantum Nambu bracket (see section 3. for a precise definition).
Also in this case
\EQ
\int d \xi^{1} d \xi^{2}... d \xi^{p+1} \rightarrow TR
\EN
where $TR$ is a suitable generalization of the operation of tracing.

Given the expressions for the light-cone and conformal gauge-fixed p-brane
actions, we
can 
obviously apply this dictionary. The classical Nambu bracket in the 
expressions for $S_{lc}$ and $S_c$ is replaced by its quantum counterpart and
integration by tracing.

The covariant p-brane action is of course also invariant 
under the subgroup of world-volume diffeomorphism. The covariant
action $S$ can be rewritten as
\EQ
S = - \int d^{p+1} \xi \sqrt{-det g_{ij}} = - 
\int d^{p+1} \xi \sqrt{{1 \over {(p+1)!}} \{x^{a_1},
x^{a_2},...,x^{a_{p+1}}\}^{2}},
\EN
or equivalently, in the polynomial form
\EQ
S = 1/2 \int d^{p+1} \xi ({1 \over (p+1)! e}\{x^{a_1},
x^{a_2},...,x^{a_{p+1}}\}^{2}
-e).
\EN
Therefore the above dictionary can be directly applied at the level 
of the covariant action (with $b$ denoting the covariant index $\mu$ and
$l=p+1$).
This procedure gives an appropriately discretized version of
the covariant p-brane action.
We will use this formal observation when we discuss the issue of 
covariantization of Matrix theory in section 4.

Note that the volume preserving diffeomorphisms
are {\it not} related to the gauge transformations of the
Yang-Mills type. In fact it was conjectured in \cite{volume}
that p-dimensional volume preserving diffeomorphisms are
related to an infinite-dimensional non-Abelian antisymmetric
tensor gauge theory.

In order to explore these
non-Yang-Mills transformations it is natural to
study the
discretization of p-branes 
by quantizing the Nambu bracket.
The solution to the quantization problem is given by the Zariski deformation 
quantization discovered in \cite{flato} which we discuss in the next section.

%%%%%%%%%%%%%%%%%%%%%%%%%%%%%%%%%%%%%%%%%%%%%%%%%%%%%

\section{Deformation Quantization and
the Nambu Bracket}

%%%%%%%%%%%%%%%%%%%%%%%%%%%%%%%%%%%%%%%%%%%%%%%%%%%%%

For simplicity, in this section we consider only the triple Nambu bracket.
Our treatment can be easily extended to the general case.

\subsection{Classical Nambu Bracket}

We start this section with a short review of the fundamental 
properties of the classical Nambu bracket following 
\cite{almy}.
Consider a three-dimensional space parametrized by
$\{x^i\}$.
The three-dimensional volume preserving diffeomorphisms (VPD) on this space
are described by a differentiable map
\EQ
x^i \rightarrow y^i (x)
\EN
such that
\EQ
\{y^1, y^2, y^3 \} =1
\EN
where, by definition, as in section 2.
\EQ
\{A, B, C\} \equiv \epsilon^{ijk}\partial_i A \partial_j B \partial_k C
\label{npbracket}
\EN
is the Nambu-Poisson bracket, or Nambu bracket, or
Nambu triple bracket, which satisfies \cite{flato},
\cite{fi}, \cite{flato2}, \cite{filip}
\begin{enumerate}
\item Skew-symmetry
\EQ
\{A_1, A_2,A_3\}=(-1)^{\epsilon(p)} \{ A_{p(1)}, A_{p(2)},A_{p(3)} \},
\label{skewsymmetry}
\EN
where $p(i)$ is the permutation of indices and
$\epsilon(p)$ is the parity of the permutation,
\item Derivation
\EQ
\{A_1A_2, A_3, A_4\} =A_1\{A_2, A_3,A_4\} + \{A_1, A_3,A_4\}A_2 ,
\EN
\item Fundamental Identity (FI-1) \cite{fi}, \cite{filip}
\EQA
\{\{A_1, A_2, A_3\},A_4, A_5 \} +\{A_3, \{A_1, A_2,A_4\},A_5\}
\nonumber \\
+\{A_3, A_4, \{A_1, A_2, A_5\}\} =
\{A_1,A_2,\{A_3, A_4, A_5\}\}.
\EQN
\end{enumerate}

The three-dimensional VPD involve two independent functions.
Let these functions be denoted by $f$ and $g$.
The infinitesimal three-dimensional VPD generator is then given as
\EQA
D{(f, g)} &\equiv& \epsilon^{ijk}\partial_i f \partial_j g \partial_k
\\
           &\equiv&  D^k(f, g) \partial_k .
\label{3dapdgenerator}
\EQN
The volume-preserving property is nothing but the identity
\EQ
\partial_i D^i(f,g) =\partial_k (\epsilon^{ijk}\partial_i f \partial_j
g )=0 .
\EN
Given an arbitrary scalar function $X(x^i)$, the three-dimensional VPD
act as
\EQ
D{(f,g)}X = \{f, g, X\} .
\EN
Apart from the issue of global definition of the functions $f$ and $g$,
we can represent an arbitrary infinitesimal volume-preserving
diffeomorphism
in this form.

On the other hand, if the base three-dimensional space $\{x^i\}$
is mapped into a target space of dimension $d+1$ whose
coordinates are $X^{\alpha} \, \, (\alpha =0,1,2, \ldots, d)$,
the induced infinitesimal volume element is
\EQ
d\sigma \equiv \sqrt{
\{X^{\alpha}, X^{\beta}, X^{\gamma}\}^2
}dx^1dx^2dx^3,
\EN
provided the target space is a flat Euclidean space.
The volume element is of course invariant under the general
three-dimensional
diffeomorphisms.

The triple product $\{X^{\alpha}, X^{\beta}, X^{\gamma}\}$ is also
"invariant" under the VPD. More precisely, it transforms as a scalar.
Namely,
\EQ
\{Y^{\alpha}, Y^{\beta}, Y^{\gamma}\} -
\{X^{\alpha}, X^{\beta}, X^{\gamma}\} =\epsilon D(f,g)\{X^{\alpha},
X^{\beta}, X^{\gamma}\} +O(\epsilon^2)
\EN
for
\EQ
Y=X+ \epsilon D(f,g)X .
\EN
This is due to the the Fundamental Identity FI-1 which shows that the
operator $D{(f,g)}$ acts as a derivation within the
Nambu bracket. For fixed $f$ and $g$, we can define a finite transformation
by \EQ
X(t) \equiv \exp (tD(f,g)) \, \rightarrow X =
\sum_{n=0}^{\infty}
{t^n\over n!}\{f, g, \{f,g, \{ \ldots , \{f,g,\{f,g, X\}\}\ldots,\}\}\}
\EN
which satisfies the Nambu ``equation of motion" \cite{nambu}
\EQ
{d\over dt}X(t)=\{f,g, X(t)\}.
\EN
The Nambu-Poisson structure is preserved under this evolution equation.

Notice that in the case of the usual Poisson structure,
the algebra of two-dimensional area preserving diffeomorphisms is given by
\EQ
[D(f_1), D(f_2)] =D(f_3)
\EN
where
\EQ
f_3= \{f_1, f_2\}
\EN
\EQ
D(f)X =\{f, X\}.
\EN
It turns out that the three-dimensional analogue
of the commutator algebra \cite{almy}
\EQ
D(A_{[1}) D(A_{2]}) =D(\{ A_1, A_2 \})
\EN
can be written using the quantum triple Nambu commutator \cite{nambu}
\EQ
[A,B,C]_N \equiv ABC-ACB+BCA-BAC+CAB-CBA
\label{nambutriple}
\EN
as follows
\EQ
D(A_{[1}, A_2) D(A_{3]_N},B) = 2D(\{ A_1, A_2, A_3 \},B),
\EN
or equivalently
\EQ
D(B_{[1}, B_2) D(A_{[1},A_2) D(A_{3]_N},B_{3]_N})=
4D(\{ A_1, A_2, A_3 \},\{ B_1, B_2, B_3 \}).
\EN
Both relations are equivalent to the Fundamental Identity
\EQA
&\{ A_1, A_2, \{A_3,B,C \}
 \} + \{ A_2, A_3, \{A_1,B,C \}
 \} + \{ A_3, A_1, \{A_2,B,C \}
 \} &
\nonumber\\
&= \{ \{A_1, A_2, A_3 \},B,C \}&.
\EQN
This result suggests that there is a new kind of
symmetry based on a new composition law
whose infinitesimal algebra is given by the
triple commutator (\ref{nambutriple}).
It was conjectured in \cite{almy} that this symmetry is  related
to the gauge transformations that are not of the Yang-Mills type
\cite{volume}.

\subsection{Deformation Quantization of the Nambu bracket}

Armed with this background we are ready to discuss the deformation quantization of the
Nambu bracket.

The basic philosophy of deformation quantization \cite{star} is to view 
quantization as the procedure of replacing the algebra of observables in 
classical mechanics,
such as functions defined over classical phase space, with the usual product 
of functions, by the algebra defined by a deformed product of functions 
(called star or Moyal product).
 
The Moyal (star) product \cite{star} of two functions on a 2-dimensional 
phase space $(q,p)$ is defined as a deformation of the ordinary product of functions
\EQ
(f * g) (q,p) =  \exp[{i\hbar/2(\partial_q \partial_p
-\partial_p \partial_q)}] (f, g) \equiv \exp[{i\hbar/2 {\cal{P}}}] (f, g)
\EN
and is known to be associative 
\EQ
(f*g)*h = f*(g*h).
\EN
The Moyal bracket is defined as
\EQ
\{ \{ f,g \}\} \equiv {1 \over i\hbar} (f *g -g*f).
\EN
The properties of the Moyal bracket are the same as the properties of
the Poisson bracket or the commutator of operators in quantum mechanics.
This enables one in principle to perform quantum-mechanical calculations within 
the classical framework without using Hilber-space formalism.
There are other formal advantages of this viewpoint.

For example, the Feyman path integral can be viewed as the Fourier 
transform over the momentum variable $p$ of 
$\exp^{*}(-itH/{\hbar})(q,p)$ \cite{star1}
where
\EQ
\exp^{*}(f) (q,p) = 1 + f(q,p) + 1/{2!} (f*f) (q,p) + ...
\EN
In the symplectic case the star product is given by the Feynman path integral. 
The general solution for the deformation quantization of the algebra of functions 
on a Poisson manifold, as formulated by Kontsevich \cite{kon}, 
can also be understood from this point of view.

It is interesting that the same set-up of Moyal deformation 
quantization does not work for the Nambu bracket.
The basic obstruction is provided by the property of the Fundamental Identity.

For example, the exponentiated Nambu bracket 
\EQ
[f_1,f_2,f_3]_{\hbar} \equiv 
{1 \over 6} \sum_{p} \epsilon(p) (f_{p1}, f_{p2}, f_{p3})_{\hbar}
\EN
where the deformed product is defined as
\EQ
(f_1,f_2,f_3)_{\hbar} = \exp[{i\hbar/2(\{\partial_x, \partial_y, \partial_z \})}]
(f_1,f_2,f_3)
\EN
{\it does not}
satisfy the property of the Fundamental 
Identity of the classical Nambu bracket \cite{flato}.

The solution of the deformation quantization problem is 
rather non-trivial in this situation and has been discovered
in \cite{flato}. The solution involves the Zariski product which is based on the
factorization of polynomials in several real variables.

Before describing the formalism we have to specify what 
we mean by a quantum triple Nambu bracket. In general we want
an object $[F,G,W]$ which satisfies the properties analogous
to the classical Nambu bracket $\{f,g,w \}$ as listed in
the previous section. (Here $f,g,w$ are
functions of three variables, and the nature of $F,G,W$ is left open
for the moment.) Thus $[F,G,W]$ is expected to
satisfy \cite{flato}, \cite{fi}, \cite{flato2}, \cite{filip}
\begin{enumerate}
\item Skew-symmetry
\EQ
[A_1, A_2,A_3]=(-1)^{\epsilon(p)} [ A_{p(1)}, A_{p(2)},A_{p(3)} ],
\label{skewsymmetry1}
\EN
where again $p(i)$ is the permutation of indices and $\epsilon(p)$
is the parity of the permutation,
\item Derivation
\EQ
[A_1A_2, A_3, A_4] =A_1[A_2, A_3,A_4] + [A_1, A_3,A_4]A_2 ,
\EN
\item Fundamental Identity (F.I.)
\EQA
[[A_1, A_2, A_3],A_4, A_5 ] &+& [A_3, [A_1, A_2,A_4],A_5]
\nonumber \\
+[A_3, A_4, [A_1, A_2, A_5]] &=&
[A_1,A_2,[A_3, A_4, A_5]].
\EQN
\end{enumerate}
(Note that the two-dimensional quantum Nambu bracket which satisfies
above properties is just the
usual commutator of matrices $[A,B] \equiv AB-BC$. In this case the
F.I. reduces to the Jacobi identity.)

Now we describe the Zariski quantization scheme following the original 
work of \cite{flato}.

The basic idea of \cite{flato} is to replace the usual product of functions in 
the classical Nambu bracket which is Abelian, associative, distributive and 
respects the Leibnitz rule, by another (deformed) product with the same properties. 
Then the modified Nambu bracket defined by such a product will 
satisfy the same properties of the classical Nambu bracket.

The construction is based on a couple of preliminary definitions \cite{flato}:

1) Let $N^{irr}$ be the set of real irreducible normalized polynomials of 
three variables $x^1, x^2, x^3$.
Let $Z_0$ be a real vector space having a basis indexed by products of elements
of $N^{irr}$. 
Denote the basis of $Z_0$ by $Z_{u_1...u_m}$ where $u_1...u_m$ are 
elements of $N^{irr}$. The vector space $Z_0$ becomes an algebra by defining a
product $\bullet$ 
\EQ
Z_{u_1...u_m} \bullet Z_{v_1...u_n} = Z_{u_1...u_m v_1...v_n}
\EN
called the Zariski product.

Let $Z_{\hbar}$ be the vector space of polynomials in 
${\hbar}$ (the deformation parameter) with coefficients in $Z_0$. Then let
\EQ
\zeta(\sum_r \hbar^r u_r) = \sum_r \hbar^r Z_{u_r}.
\EN
The deformed Zariski product is defined as
\EQ
Z_{u_1...u_m} \bullet_{\hbar} Z_{v_1...u_n} = 
\zeta((u_1,...u_m) \times_{\alpha} (v_1...v_n))
\EN
where the operation $\times_{\alpha}$ for two 
irreducible polynomials $u_1$ and $v_1$ is given by the folowing formula 
\EQ
u_1 \times_{\alpha} v_1 = {1 \over 2} (u_1 * v_1 + v_1*u_1)
\EN
and where $*$ is the usual Moyal product with respect to two variables $x_1$ and
$x_2$.
In general, one has to fully symmetrize the $*$ product of irreducible
polynomials. Also, the deformed product is extended from $Z_0$ to
$Z_{\hbar}$ by requiring that it annihilates the non-zero
powers
of the deformation parameter.

2) Let $E = Z_0[y^1, y^2, y^3]$ be the algebra of polynomials 
in three variables $y^1, y^2, y^3$ with coefficients in $Z_0$.
Let $A_0$ be a
subalgebra of $E$ generated by the following "Taylor series"
in $E$
\EQ
J(Z_u) = Z_u + \sum_i y^i Z_{\partial_{i} u} +
{1 \over 2} \sum_{i,j} y^i y^j Z_{\partial_{ij} u} +... \equiv \sum_n
{ 1 \over n!} (\sum_i y^i \partial_i)^{n} (Z_{u})
\EN
where $\partial_i Z_u \equiv Z_{\partial_i u}$ and 
$u$ are normalized polynomials in three variables $x^1, x^2, x^3$ 
that can be uniquely factorized in $u_1,...,u_m$, and $\partial_i$ is a
derivative
with respect to $x^i$.
Define the following generalized derivative
\EQ
\Delta_{a}(J(Z_u)) \equiv J(Z_{\partial_a u})
\EN
where $a=1,2,3$.

Then the classical Nambu bracket on $A_0$ is defined as \cite{flato}
\EQ
[A,B,C]_{\bullet} = \sum_{p} \epsilon(p) \Delta_{p_1} A \bullet
\Delta_{p_2} B \bullet
\Delta_{p_3} C
\EN
for three elements $A,B,C$ from $A_0$.

It can be shown \cite{flato} that there exists a deformation of this Nambu bracket
in the sense of deformation quantization.

Let $E_{\hbar}$ be the algebra of polynomials of $\hbar$ (the deformation
parameter) with coefficients in $E$ defined above.
Consider a subspace $A_{\hbar}$ of $E_{\hbar}$ consisting of polynomials in
$\hbar$ with coefficients in $A_0$.
Then the authors of \cite{flato} define a deformed product
\EQ
J(Z_u) \bullet_{\hbar} J(Z_v)
= Z_u \bullet_{\hbar} Z_v + 
\sum_i y^i (Z_{\partial_{i} u} \bullet_{\hbar} Z_u + 
Z_u \bullet_{\hbar} Z_{\partial_{i} u}Z_u )+
...
\EN 
where the deformed Zariski product of basis elements is defined
as before.

Then one can define the quantum Nambu bracket on $A_{\hbar}$ as follows
\EQ
[A,B,C]_{\bullet_{\hbar}} = \sum_{p} \epsilon(p) \Delta_{p_1} A \bullet_{\hbar}
\Delta_{p_2} B \bullet_{\hbar}
\Delta_{p_3} C
\EN
for three elements $A,B,C$ from $A_{\hbar}$.
 
It can be shown \cite{flato} that this quantum Nambu bracket satisfies
all three fundamental properties: 
it is totally antisymmetric, it satisfies the 
derivation property and the Fundamental Identity.

Therefore, this construction solves the quantization problem.

Obviously the same construction can be applied to the usual Poisson bracket.
In that case the Zariski quantized 
Poisson bracket is {\it not} the skew-symmetrized 
form of an associative (star) product.

Nevertheless, one can define a Moyal-Zariski product by replacing \cite{flato}
\EQ
\exp[{i\hbar/2 {\cal{P}}}]  \rightarrow
\exp[{i\hbar/2 {{\cal{P}}_{\bullet_{\hbar}}}}] 
\EN
in the defining formula for the Moyal bracket.
This procedure gives another associative deformation of the usual product
of functions. The skew-symmetrized form of this associative deformed product
will provide a Lie algebra deformation.
Note that the leading term in the
expansion of such a bracket
is given by the Zariski quantized Poisson bracket!

This fact is of crucial importance for our proposal concerning the link
between covariant Matrix theory and Zariski deformation quantization (see next
section).

We conclude this section with the following comment:

The star product relevant for the general deformation quantization 
of the Poisson bracket, as discovered by Kontsevich, can be understood from the
path integral point of view \cite{kon}.
It is natural to ask whether the star product 
(such as the deformed Zariski product) relevant for the quantization
of the triple Nambu bracket can be understood from the path integral point of view.
One obvious idea that comes to mind is that such a path 
integral has to be over loop variables, which appear naturally
in the Hamiltonian formulation of the Nambu mechanics \cite{fi}.

%%%%%%%%%%%%%%%%%%%%%%%%%%%%%%%%%%%%%%%%%%%%%%%%%%%%%%%%
\section{Covariant Matrix Theory and Zariski Quantization}
%%%%%%%%%%%%%%%%%%%%%%%%%%%%%%%%%%%%%%%%%%%%%%%%%%%%%%%%%

In this section we propose that the problem of covariantization of Matrix theory
can be solved using
the formalism of Zariski quantization.

Recall that Matrix theory \cite{matrix} uses $N\times N$ hermitian matrices 
to represent the transverse coordinates (and their super-partners) of $N$
$D0$-branes 
in the (discretized) light-cone frame 
$X^i  \quad ( i=1,2, \ldots, 9)$. The statistics of 
$D0$-branes are encoded in the 
$U(N)$ gauge symmetry ($t=X^{+}=$light-cone time) 
$X^i \rightarrow U X^i U^{-1}$.
The number $N$ of $D0$-branes is connected 
to the longitudinal momentum $P^{+}$ in the light-like direction by 
$P^{+} = {N \over R}$
where $R=g_s\ell_s$ is the compactified radius in the $X^{-}$  
direction. Also $\alpha' = l_{p}^{3}/R$, where $l_{p}$ is the
11-dimensional Planck length.

What should be the structure of a covariant Matrix theory?
We expect at least 
\begin{enumerate}
\item Generalization of the matrix algebra and 
the emergence of higher symmetry
\item $R$ (and perhaps $N$) should appear as dynamical variables. 
\end{enumerate}
At  present, the only guiding principle in trying to search for
such a formulation  
is that the formalism should reduce in the 
light-cone gauge to Matrix theory after appropriately 
fixing the gauge using the higher symmetry, 
and that it should have 11-dimensional (super) Poincare invariance 
in the limit $R\rightarrow \infty$. 

The light-cone degrees of freedom (D0-branes) carry the quantum number of 
11-dimensional (super)gravitons. Thus we expect that the covariant
version of Matrix theory will describe a many-body quantum theory of
interacting 11-dimensional (super)gravitons.

The covariant theory is also expected to be a quantum mechanical 
theory invariant under world-line reparametrizations, 
simply because in the light-cone gauge 
we have a quantum mechanical theory with a globally defined time.

Of course it is tempting to try to quantize the 
world-volume membrane theory for which we 
know a classical covariant action principle in view of the
results in sections 2. In particular, as we have explicitly said
in section 2.,
the expression for the world volume of a membrane is invariant under the
classical 3D volume preserving diffeomorphisms. 
Thus it is natural to expect that a discretized membrane 
theory should be formulated in terms of discretized volume 
preserving diffeomorphisms. (A preliminary attempt at 
discretization of
the world-volume membrane theory was reported in \cite{hadm}.)

This idea appears natural as a generalization of the  following
pictorial correspondence 

{\large
\begin{center}
\begin{tabular}{ccc}
matrices/commutators &$\leftrightarrow$& 2D surface \\
U(N) &$\leftrightarrow$& 2D APD \\
\end{tabular}

\[\Downarrow
\]
\begin{tabular}{ccc}
triple Nambu bracket &$\leftrightarrow$& 3D volume \\
? &$\leftrightarrow$& 3D diffeos \\
\end{tabular}
\end{center}}

\noindent
since the two indices of the square matrices in Matrix theory just 
correspond to the discretized  Fourier indices on 
the membrane. 
  
In the following we propose a precise algebraic structure that replaces 
the question mark in the above diagram.

%%%%%%%%%%%%%%%%%%%%%%%%%%%%%%%%%%%%%%%%%%%%%%%%%%%%%%%%
\subsection{The Proposal}
%%%%%%%%%%%%%%%%%%%%%%%%%%%%%%%%%%%%%%%%%%%%%%%%%%%%%%%%%

Here we state a concrete proposal for the covariantization of Matrix theory.
For simplicity, we discuss the bosonic part only.
Our proposal can be extended to include maximal supersymmetry.

We propose that the following action defines a covariant Matrix theory
\EQ
S_M = - \int d^{3} \xi \sqrt{{1 \over 6} [X^{\mu},
X^{\nu},X^{\rho}]^{2}_{\bullet_{1 \over N}}}.
\EN
Here $\mu = 0,1,...,10$ and $X^{\mu}$ are elements of
$A_{{1 \over N}}$ as defined in section 3. - in other words they are 
polynomials in $1/N$ with coefficients in $A_0$ 
(recall that $A_0$ is the algebra of "shifted" polynomials of 
three variables, in this case,
$\xi_1, \xi_2, \xi_3$). We take $1/N$ to be the deformation parameter of
the Zariski
quantization.

$N$ is interpreted as the number of 11-dimensional (super)gravitons.

The action $S_M$ is invariant under Zariski deformed world-volume diffeomorphisms
$\delta X^{\mu} = \epsilon^{i} \bullet_{1 \over N} \Delta_{i} X^{\mu}$,
with $\Delta$ from section 3.

Due to the usual properties of the product $\bullet_{1 \over N}$  
(recall that the deformed product is Abelian, associative and distributive and the
formal derivative $\Delta_a$ respects linearity, the Leibnitz rule
and the commutativity of the derivatives in many variables) 
we can formally repeat the same steps of fixing the light-cone gauge
for this theory, as in section 2.

Thus, the light cone action is given by a Zariski deformed 
version of the usual-light cone action for the 
M-theory membrane, with a deformation parameter
$1/N$
\EQ
S_{lc}= 1/2 \int d^{3} \xi ( (\Delta_0 X^{a})^{2} - 
(1/2) [X^{a}, X^{b}]^{2}_{\bullet_{1 \over N}} ). 
\EN
Now we use the crucial observation from section 3.
The skew-symmetrized form of the Moyal-Zariski associative deformed product
provides a Lie algebra deformation, and is thus in 
the same class as the commutator of matrices. Moreover, 
the leading term (in $N$) in the expansion of such a bracket
is given by the Zariski quantized Poisson bracket.

In view of this fact, it is natural to propose that this 
Zariski deformed light-cone action is in the same
universality class as the Matrix theory action \cite{matrix}, where 
$N$ is the number of D0-branes.

Thus the above covariant action can be viewed as a description of a 
large $N$ quantum mechanical theory of $N$ interacting 
11-dimensional (super)gravitons.

Notice that our proposal can be formulated for the Polyakov type 
action as well, due to the properties of the deformed Zariski product.
Because of the usual properties of the deformed product 
$\bullet_{1 \over N}$ it is possible to formulate our proposal 
for the supersymmetric case as well. 

Finally we comment on the $N \rightarrow \infty$ limit of our proposal.

The $N \rightarrow \infty$ of the action $S_M$ reduces to the bosonic 
part of the usual M-theory membrane action with space-time 
coordinates $x^{\mu}$ being the elements of the algebra $A_0$,
where $\bullet$ defines the two product and 
$\Delta$ the formal derivation with respect to the world-volume coordinates
$\xi$.
If our proposal is correct this $N \rightarrow \infty$ limit theory should be in
the
same universality class as the ordinary bosonic membrane \cite{membrane}.
The same comments apply to the supersymmetric case.

As in Matrix theory, the membrane is just one sector of the theory.
Other physical sectors should exist as well.

We conclude this subsection with a list of a few obvious questions:

- is there an explicit matrix version of our conjecture ?

- what is the physical interpretation of anti-branes in our proposal?

- does our proposal include the M-theory five-brane in the large N limit?

At present we do not have definitive answers to these important questions, 
so we postpone their discussion for the future.

%%%%%%%%%%%%%%%%%%%%%%%%%%%%%%%%%%%%%%%%%%%%%%%%%%%%%

\section{Discussion}

%%%%%%%%%%%%%%%%%%%%%%%%%%%%%%%%%%%%%%%%%%%%%%%%%%%%%

In this, concluding section, we compare the construction of quantum Nambu
brackets in terms of square and cubic matrices \cite{almy} to the Zariski 
deformation quantization scheme and discuss a natural mathematical formulation of 
the space-time uncertainty principle of M-theory \cite{stu} in view of our proposal.

%%%%%%%%%%%%%%%%%%%%%%%%%%%%%%%%%%%%%%%%%%%%%%%%%%%%%

\subsection{Square and Cubic Matrices vs. Zariski Quantization}

%%%%%%%%%%%%%%%%%%%%%%%%%%%%%%%%%%%%%%%%%%%%%%%%%%%%%

An explicit
matrix realization of the
quantum Nambu bracket, which is skew-symmetric and obeys the
Fundamental Identity was given in \cite{almy}. This realization was contrasted to
the example constructed in \cite{flato2}.

Define a totally antisymmetric
triple bracket of three matrices A, B, C as
\EQ
[A,B,C] \equiv
({\rm tr} A)  [B,C] +(tr B)[C,A]+(tr C)[A,B] \label{31}.
\EN
Then ${\rm tr} [A,B,C]=0$, and if $C=1$, $[A,B,1]=N [A,B]$, where
$N$ is the rank of square matrices. 
This bracket is obviously skew-symmetric and it can be shown to 
obey the Fundamental Identity \cite{almy}.

Given this example of a three-dimensional quantum Nambu bracket, 
consider the following "gauge transformation"
\EQ
{\delta} A
\equiv i[X,Y,A], \label{32}
\EN
where the factor $i$ is introduced for
Hermitian matrices. This transformation represents an obvious quantum form of the
three-dimensional volume preserving diffeomorphisms.
By the definition of
the triple bracket, the generalized gauge transformation takes the following
explicit form
\EQ
\delta A=i\left( [({\rm tr} X)Y-({\rm tr} Y)X, A]+({\rm tr} A)[X,Y]\right).
\label{33}
\EN
It is shown in \cite{almy} that if ${\rm tr} A_i=0, i=1, ... n$, then
\EQ
{\rm tr} (A_1A_2\dots A_n)
\EN
is gauge invariant.
Note that the gauge transformation of
a commutator does not satisfy the usual composition rule, namely
$$
[X,Y, [A,B]]\ne [[X,Y,A],B]+[A,[X,Y,B]] \label{34}.
$$
Similar comments apply to $[[X^i,X^j],[X^l,X^k]]$.

Notice that the form of the gauge transformation (\ref{33})
indicates that a bosonic
Hermitian matrix $A$  can be transformed into a form proportional to
the unit $N \times N$ matrix as long as ${\rm tr} A\ne 0$. In other
words, since the gauge transformation is traceless,
one can show that a Hermitian matrix can be brought to the following form
$$A\rightarrow {1\over N}{\rm tr} A 1_N.$$

This matrix realization of
the triple quantum Nambu bracket can be generalized to a representation in
terms of
three-index objects or cubic matrices \cite{almy}.

To this end, introduce the following generalization of the trace
\EQ
\< A \> \equiv \sum_{pm} A_{pmp},\qquad
\< A B \> \equiv \sum_{pqm} A_{pmq} B_{qmp},\qquad
\< ABC \> \equiv \sum_{pqrm} A_{pmq} B_{qmr} C_{rmp},
\label{e:trace}
\EN
which satisfy $\<AB\>=\<BA\>$ and $\<ABC\>=\<BCA\>=\<CAB\>$.
Furthermore, define a triple-product
\begin{equation}
(ABC)_{ijk} \equiv \sum_{p} A_{ijp} \< B \> C_{pjk}
=\sum_{pqm} A_{ijp} B_{qmq} C_{pjk}. \label{t11}
\end{equation}
and the following skew-symmetric
quantum Nambu bracket
\begin{equation}
[A,B,C] \equiv
  (ABC) + (BCA) + (CAB)
- (CBA) - (ACB) - (BAC).
\label{e:NambuBracket}
\end{equation}
The middle index $j$ of $A_{ijk}$ can be treated
as an internal index for the matrix realization of the
triple quantum Nambu bracket. Note also that
$\<(ABC)\> = \<B\>\<AC\> \neq \<ABC\>$ and
$\<(ABC)D\> = \<B\>\<ACD\>$.

Then by using the following relations
\EQA
((ABC)DE) &=& ((ADC)BE) = (AB(CDE)) = (AD(CBE)),\cr
(A(BCD)E) &=& (A(DCB)E), \label{t12}
\EQN
%$(A[B,C,D]E) = 0$ and
one can directly prove that
the skew-symmetric Nambu bracket $(\ref{e:NambuBracket})$
with the triple-product $(\ref{t11})$ obeys the Fundamental Identity \cite{almy}.
%$$[X,Y,[A,B,C]] = [[X,Y,A],B,C] + [A,[X,Y,B],C] + [A,B,[X,Y,C]].$$

The ``trace'' $\<AB\>$ has the property
$$\<\,[X,Y,A] B\,\> + \<\,A [X,Y,B]\,\> = 0,$$ provided $\<A\> = \<B\> = 0$.
Therefore, since $\<\,[A,B,C]\,\> = 0$ for any three-index 
objects $A$, $B$ and $C$,
the trace of the product of Nambu brackets
$\<\,[A,B,C] [D,E,F]\,\>$ is gauge invariant.
Notice that if one generalizes the trace (\ref{e:trace}) as
\EQ
\< A^1 A^2 \cdots A^n \> \equiv \sum_{p_1,p_2,\cdots,m}
A^1_{p_1 m p_2} A^2_{p_2 m p_3} \cdots A^n_{p_n m p_1},\qquad n=1,2,\cdots,
\EN
one can also demonstrate that the trace of any product of Nambu brackets
$\<\,[A,B,C] [D,E,F] \cdots [X,Y,Z]\,\>$ is gauge invariant \cite{almy}.
%which enjoys $\<A^1 A^2 \cdots A^n\>=\<A^2 \cdots A^n A^1\>$.
%This has the property
%\EQ  \< [X,Y,A^1] A^2 \cdots A^n \> + \< [X,Y,A^2] A^3 \cdots A^n A^1 \>
%\cdots + \< [X,Y,A^n] A^1 \cdots A^{n-1} \> = 0,\EN provided $\<A^i\> = 0$.
%Therefore, since $\<[A,B,C]\> = 0$,
%{\it the trace $\<\cdots\>$ of any product of the quantum Nambu brackets
%$\<[A,B,C] [D,E,F] \cdots [X,Y,Z]\>$ is gauge invariant.}

%\noindent
%{\bf  Identity Element.}
Define
$I_{ijk} \equiv \delta_{ik}^{(j)}$,
where $\delta_{ik}^{(j)} =0$, if $i \neq k$, for any $j$ and
$\delta_{ik}^{(j)} =1$, if $i = k$, for any $j$.
Then
\EQA
(AIB) = \<I\> \sum_p A_{ijp} B_{pjk},&&
(IAB) = (BAI) = \<A\>B,
\cr
(IAI) = \<A\> \delta_{ik}^{(j)},\qquad &&
(IIA) = (AII) = \<I\> A,
\EQN
and
$[A,I,B] = \sum_p (A_{ijp} B_{pjk} - B_{ijp} A_{pjk})$.
Hence for any middle index $j$,
$[A,I,B]$ reduces to the usual commutator $[A^{(j)},B^{(j)}]$
for the matrices
$A^{(j)}_{ik} \equiv A_{ijk}$ and
$B^{(j)}_{ik} \equiv B_{ijk}$.

Other examples of triple-products $(ABC)_{ijk}$
which also satisfy the same relations as eq.\ (\ref{t12})
and hence lead to the F.I.
for the skew-symmetric Nambu bracket (\ref{e:NambuBracket}) were listed in
\cite{almy}
%\EQA
%(ABC)_{ijk} = \sum_{pq} A_{ijp} B_{qjq} C_{pjk},\cr
%(ABC)_{ijk} = \sum_{pqmn} A_{ijp} B_{qmq} C_{pnk},\cr
%(ABC)_{ijk} = \sum_{pqmn} A_{inp} B_{qmq} C_{pjk},\EQN
%\EQA
%(ABC)_{ijk} &=& \sum_{pq} A_{ijp} B_{qjq} C_{pjk},\cr
%(ABC)_{ijk} = \sum_{pqmn} A_{ijp} B_{qmq} C_{pnk},&&
%(ABC)_{ijk} = \sum_{pqmn} A_{inp} B_{qmq} C_{pjk},\EQN
\EQ
\sum_{pq} A_{ijp} B_{qjq} C_{pjk},\qquad
\sum_{pqmn} A_{ijp} B_{qmq} C_{pnk},\qquad
\sum_{pqmn} A_{inp} B_{qmq} C_{pjk}.
\EN

The explicit examples of the quantum Nambu bracket 
presented above do not satisfy the derivation property. 
Also, the form of the above quantum Nambu brackets does not completely parallel
the form
of the classical Nambu bracket.

The classical triple Nambu bracket of three functions $f, g, h$ of three 
variables $\tau,\sigma_1, \sigma_2$ can be obviously rewritten as
\EQ
\{ f, g, w \} = \dot{f} \{g,w\} + \dot{g} \{w,f\} + \dot{w} \{f,g\}
\EN
where $\dot{f}= \partial_{\tau} f$ and
$\{f,g\} = \partial_{\sigma_1} f \partial_{\sigma_2} g
- \partial_{\sigma_2} f \partial_{\sigma_1} g$.

If we try to extrapolate our previous definition of the 
quantum Nambu bracket in terms of $\tau$-dependent square matrices to
\EQ
[ F, G, W ] = \dot{F} [G, W] + \dot{G} [W, F] + \dot{W} [F, G]
\EN
where $\dot{F}= \partial_{\tau} F$ and the usual matrix multiplication is assumed,
one can show that such $[F, G, W]$ does not satisfy the Fundamental Identity
\footnote{This was explicitly shown by H. Awata - private communication.}. 
From the analogy with the
covariant membrane and our discussion in section 4., we interpret $\tau$ 
as a potential world-line parameter.

Obviously the Zariski deformation quantization provides 
us with a quantum Nambu bracket which has the 
same properties as its classical counterpart and 
therefore can be used for the discretization of $p$-brane actions as 
discussed in subsection 2.3.
That is why our proposal toward a covariant formulation of Matrix theory from
section 4. was formulated in terms of the algebraic structures needed in
the Zariski
deformation quantization.

It is not clear at present whether  
the Zariski deformation quantization has an explicit 
representation in terms of matrices (either square or cubic).

%%%%%%%%%%%%%%%%%%%%%%%%%%%%%%%%%%%%%%%%%%%%%%%%%%%%%%%%
\subsection{Space-Time Uncertainty Relation in M-theory}
%%%%%%%%%%%%%%%%%%%%%%%%%%%%%%%%%%%%%%%%%%%%%%%%%%%%%%%%%

The main signature of the space-time uncertainty relation 
\cite{stu} is 
the opposite scaling of the transverse and longitudinal directions
with respect to the fundamental string (in perturbative string theory) or
with respect to a $Dp$-brane  in non-perturbative string/M
theory (see also \cite{uvir}). 
In perturbative string theory the space-time 
uncertainty relation incorporates the effects of conformal symmetry. 
In non-perturbative string/M theory, the space-time uncertainty 
relation captures the essential features of the physics of D-branes 
as well as the property of holographic behavior \cite{hol}.
Thus it is reasonable to expect that the space-time uncertainty 
relation says something important 
about the underlying physical foundation of M-theory. 
Here we want to comment on the mathematical 
structure underlying the space-time uncertainty relation in view of our proposal
from section 4.

Perhaps the simplest way of characterizing the space-time uncertainty
relation is by writing
\EQ
\delta T \delta X \sim \alpha' , \label{str}
\EN
where $\delta T$ and $\delta X$ measure the effective longitudinal (along
the world-volume of a p-brane)
and transverse space-time distances. 
This equation is true in Matrix theory \cite{matrix}
\EQ
X^{a} \rightarrow \lambda X^{a}, t \rightarrow {\lambda}^{-1} t, \label{xt}
\EN
provided the longitudinal distance
is identified with the global time of Matrix theory. 
The effective super Yang-Mills theory which describes the dynamics of $D0$-branes 
is
invariant under (\ref{xt}) if the string coupling constant is simultaneously rescaled
\EQ
g_s \rightarrow {\lambda}^{3} g_s .
\EN
The space-time uncertainty relation leads readily to the well known
characteristic space-time scales in M-theory \cite{stu}.
Note that eq. (\ref{xt}) can also be applied to $Dp$-branes, provided
\EQ
g_s \rightarrow {\lambda}^{3-p} g_s .
\EN

What is the natural mathematical set-up for the 
space-time uncertainty relation in string theory (\ref{str})?
It was proposed by Yoneya and Li and Yoneya \cite{stu}
that the right point of view for discussing 
the space-time uncertainty principle 
in string theory, is to treat all space-time coordinates
as infinite dimensional matrices.

Remarkably, the string theory space-time uncertainty relation can be
understood as a limit of the space-time uncertainty relation in M-theory as
noticed by Li and Yoneya \cite{stu}. 
In Matrix theory eq.(\ref{str}) can be rewritten as
\EQ
\delta T \delta X_{T} \sim l_{p}^{3}/R , \label{mst}
\EN
where $\delta X_{T}$ and $\delta T$ 
respectively measure transverse spatial and time directions.

Li and Yoneya have proposed a more general relation \cite{stu} by observing
that the uncertainty for the longitudinal direction in physical
processes that involve individual $D0$-branes is 
$\delta X_{L} \sim R$. Then (\ref{mst}) reads
\EQ
\delta T \delta X_{T} \delta X_{L} \sim l_{p}^{3} . \label{mst1}
\EN
This is the space-time uncertainty relation in M-theory. Note
that only the fundamental length scale of M-theory ($l_{p}$)
figures in this relation.

The point we want to make is that the form of the 
space-time uncertainty relation in M-theory is very reminiscent of the form of the
quantum triple Nambu bracket we have discussed in the previous
section \cite{almy}.
In particular it is natural to extend the proposal of Yoneya
and Li and Yoneya \cite{stu}
for the case of Matrix theory, and suggest a covariant version of
the space-time uncertainty relation in Matrix theory which is consistent
with (\ref{mst1})
\EQ
[X^{\mu}, X^{\nu}, X^{\lambda}]^{2}_{\bullet_{1 \over N}} \sim l_{p}^{6}
\EN
Here $\mu, \nu, \lambda = 0,1,...,10$.
The triple bracket in the above formula is the quantum triple Nambu
bracket in the sense of Zariski deformation quantization with the 
deformation parameter ${1 \over N}$ and $X^{\mu}$'s are elements of
$A_{{1 \over N}}$ as in the previous section.

In view of the relation between 
the space-time uncertainty principle  
\cite{stu} and the UV/IR relation \cite{uvir} and the holographic principle 
\cite{hol}
it is natural to expect that this mathematical formulation 
of the space-time uncertainty principle captures the 
mathematical content of the holographic principle in the flat space-time limit.

%%%%%%%%%%%%%%%%%%%%%%%%%%%%%%%%%%%%%%%%%%%%%%%%%%%%%

\vspace{1cm}
\noindent
{\bf Acknowledgements}

%%%%%%%%%%%%%%%%%%%%%%%%%%%%%%%%%%%%%%%%%%%%%%%%%%%%%

It is my pleasure to thank  H. Awata, I. Bars, S. Chaudhuri,
M. G\"{u}naydin, P. Ho\v{r}ava, T. H\"{u}bsch, M. Li, J. Polchinski,
K. Pilch, J. Schwarz,
L. Smolin, E. Witten, C. Zachos and T. Yoneya for interesting comments and
discussions on the problem of a covariant formulation of Matrix theory. 
Special thanks to H. Awata, M. Li and T. Yoneya 
for an enjoyable collaboration on the problem of quantization of the Nambu
bracket. 
This work is supported in part by DOE grant
DE-FG03-84ER40168 and by a National Science Foundation grant
NSF9724831 for collaborative research between USC and Japan.

%%%%%%%%%%%%%%%%%%%%%%%%%%%%%%%%%%%%%%%%%%%%%%%%


\begin{thebibliography}{99}

%%%%%%%%%%%%%%%%%%%%%%%%%%%%%%%%%%%%%%%%%%%%%%%

\bibitem{matrix} T. Banks, W. Fischler, S. H. Shenker and L. Susskind,
Phys. Rev. {\bf D55} (1997) 5112.
%
\bibitem{ads} J. Maldacena, Adv. Theor. Math. Phys, {\bf 2} (1998) 231.
%
\bibitem{pol} J. Polchinski, hep-th/9903165 and references therein; A. Jevicki and T. Yoneya, hep-th/9805069; I. Chepelev, hep-th/9901033.
%
\bibitem{nonc} A. Connes, M. R. Douglas and A. Schwarz, JHEP, {\bf 9802:003}
(1998); M. R. Douglas and C. Hull, JHEP {\bf 9802:008} (1998);
for reviews and further references consult, M. R. Douglas, hep-th/9901146;
N. Seiberg and E. Witten, hep-th/9908142.
%
\bibitem{flato} G. Dito, M. Flato, D. Sternheimer and L. Takhtajan, Comm.
Math. Phys. {\bf 183} (1997) 1.
%
\bibitem{volume} E. Bergshoeff, E. Sezgin, Y. Tanii and P. K. Townsend,
Ann. Phys. {\bf 199} (1990) 340.
%
\bibitem{membrane} E. Bergshoeff, E. Sezgin and P. K. Townsend,
Phys. Lett. {\bf 189B} (1987) 75;
Ann. Phys. {\bf 185} (1988) 330;
E. Bergshoeff, E. Sezgin and Y. Tanii, Nucl. Phys. {\bf B298} (1988) 187;
B. de Wit, J. Hoppe and H. Nicolai, Nucl. Phys. {\bf B305} (1988) 545;
I. Bars, hep-th/9706177.
%
\bibitem{area} J. Goldstone, unpublished;
J. Hoppe, MIT Ph.D. thesis, 1982 and in
"Proc. Int. Workshop on Constraint's Theory and Relativistic Dynamics",
G. Longhi and L. Lusanna, eds. (World Scientific, 1987);
J. Hoppe, Int. J. Mod. Phys. {\bf A4} (1989) 5235;
D. Fairlie, P. Fletcher and C. Zachos, J. Math. Phys. {\bf 31} (1990) 1088.
%
\bibitem{bars} I. Bars, Nucl. Phys. {\bf B343} (1990) 398.
\bibitem{nambu} Y. Nambu, Phys. Rev. {\bf D7} (1973) 2405.
\bibitem{flato1} F. Bayen and M. Flato, Phys. Rev. {\bf D11} (1975) 3049.
\bibitem{fi} L. Takhtajan, Comm. Math. Phys. {\bf 160} (1994) 295.
%
\bibitem{almy} H. Awata, M. Li, D. Minic and T. Yoneya,
hep-th/9906248.
%
\bibitem{stu} T. Yoneya, Mod. Phys. Lett. {\bf A4} (1989) 1587.
See also T. Yoneya, in "Wandering in the Fields".
K. Kawarabayashi and A. Ukawa, eds. (World Scientific, 1987) pp.419;
and T. Yoneya, in "Quantum String Theory",
N. Kawamoto and T. Kugo, eds. (Springer, 1988) pp.23;
T. Yoneya, Prog. Thor. Phys. {\bf 97} (1997) 949;
M. Li and T. Yoneya, Phys. Rev. Lett. {\bf 78} (1997) 1219;
M. Li and T. Yoneya, hep-th/9806240;
D. Minic, Phys. Lett. {\bf B442} (1998) 102; C. S. Chu, P. M. Ho and Y. C. Kao, hep-th/9904133.
%
\bibitem{hoppe} J. Hoppe, Helv. Phys. Acta {\bf 70} (1997) 302.
%
\bibitem{flato2} G. Dito and M. Flato, Lett. Math. Phys. {\bf 39} (1997)  
107.
%
\bibitem{filip} V. T. Filipov, Sib. Math. Jour. {\bf 26} no. 6 (1985) 126;
M. Flato and C. Fronsdal, unpublished.
%
\bibitem{star} F. Bayen, M. Flato, C. Fronsdal, A. Lichnerowicz and
D. Sternheimer, Ann. Phys. {\bf  111} (1978) 61-110 and 111-151.
%
\bibitem{star1} P. Sharan, Phys. Rev. {\bf D20} (1979) 414.
%
\bibitem{kon} M. Kontsevich, q-alg/9709040;
For a path integral interpretation of this results see
A. S. Cattaneo and G. Felder, math.qa/9902090.
%
\bibitem{hadm} H. Awata and D. Minic, JHEP04 (1998) 006;
K. Fujikawa and K. Okuyama, Phys. Lett. {\bf B411} (1997) 261;
L. Smolin, Phys. Rev. {\bf D57} (1998) 6216.
%
\bibitem{uvir} L. Susskind and E. Witten, hep-th/9805114; 
A. W. Peet and J. Polchinski, hep-th/9809022.
%
\bibitem{hol} G. 't Hooft, Class. Quant. Grav. {\bf 11} (1994) 621,
gr-qc/9310006;
L. Susskind, J. Math. Phys. {\bf 36} (1995) 6377, hep-th/9409089.
%


\end{thebibliography}
\end{document}